# Impedance Spectroscopy Study in the vicinity of Ferroelectric Phase Transition

**Hitesh Borkar[1], M Tomar[2], Vinay Gupta[3], Ashok Kumar[1],∗**

[1]CSIR-National Physical Laboratory, Dr. K. S. Krishnan Marg, New Delhi 110012, India
[2]Department of Physics, Miranda House, University of Delhi, Delhi 110007, India
[3]Department of Physics and Astrophysics, University of Delhi, Delhi 110007, India

An impedance spectroscopy (IS) is a versatile tool to study the effect of grains (bulk), grain boundaries and electrode-electrolyte interface on dielectric and electrical properties of electro-ceramics. This study only focuses the high frequency (1 kHz to 10 MHz) probe of bulk ferroelectric capacitance near ferroelectric phase transition temperature (FPTT). The $Pb[(Zr_{0.52}Ti_{0.48})_{0.60}(Fe_{0.67}W_{0.33})_{0.40}]O_3$ (PZTFW) single phase and $[Pb[(Zr_{0.52}Ti_{0.48})_{0.60}(Fe_{0.67}W_{0.33})_{0.40}]O_3]_{0.80}–[CoFe_2O_4]_{0.20}$ (PZTFW-CFO) composites, respectively, have been investigated to understand the microstructure-property relation. Keeping in mind the complex microstructure of both systems, low frequency (< 1 kHZ) impedance investigation, which basically deals with grain boundaries and electrode-interfaces, has been ignored. X-ray diffraction (XRD) patterns, microstructures, dielectric spectra, and impedance plots revealed two distinct phases, inbuilt compressive strain, small shift in dielectric maximum temperature, and two activation energy regions, respectively, in PZTFW-CFO composite compare to PZTFW ceramic. Addition of CFO in PZTFW medium purified the impurity phases present in the PZTFW matrix. The PZTFW-CFO composite shows flat dielectric behavior and high dielectric constant near FPTT at high frequency may be useful for tunable dielectric capacitors. The changes in bulk capacitance, relaxation time and constant phase element parameters have probed in the proximity of FPTT regions. Nyquist plot and modulus formalism show a poly-dispersive nature of relaxation, relate the activation energy ($E_a$) of oxygen vacancies, mainly responsible for the bulk capacitive conduction. A spiral kind of modulus spectra was observed at elevated temperatures and frequencies (> 2 MHz) suggests the possible experimental artifacts, have no physical reasons to explain.

Corresponding Author: Dr. Ashok Kumar: (Email: ashok553@nplindia.org)

I. Introduction

Since the advent of scientific interest on ferroelectric (FE) and dielectric materials, large number of researchers around the globe, are searching for a novel materials having large dielectric constant, high polarization, low coercive field, near room temperature FPTT, due to plenty of applications in microelectronics, actuators and sensors industries.[1,2] $Pb(Zr_{1-x}Ti_xO_3)$ is one such promising candidates which possesses above motioned properties, however, lowering its FE phase transition temperature near the room temperature with intact functional properties is still a challenges. [3,4] The $Pb(Zr_{0.53}Ti_{0.47})O_3$ (PZT) near morphotropic phase boundaries (MPB) is well known classical ferroelectric with phase transition temperature ($T_c$) = 620 K [5] and $Pb(Fe_{0.67}W_{0.33})O_3$ is relaxor ferroelectrics with $T_m$ = 180 K and antiferromagnetic at $T_N$ = 343 K. [6,7] Solid-solutions of these two ferroelectrics yield many more ferroelectrics and multiferroics near the room temperature. [8]

Since the re-discovery of single phase $BiFeO_3$, [9] and fabrication of controlled $BiFeO_3$-$CoFe_2O_4$ composites [10], respectively, change the direction of oxide research along multiferroic (higher ferroic order parameters). PZTFW complex perovskite has $ABO_3$ structure, with A-site occupied by $Pb^{2+}$ cation and B-site by complex compositions of $Zr^{4+}$, $Ti^{4+}$, $Fe^{3+}$, $W^{3+}$(transition metals). Discovery of multiferroic and high magnetoelectric coefficients at room temperature in these systems with proper compositional tuning of $Fe^{3+}/Ta^{5+}/Nb^{5+}$ cations at B-site, make it potential candidate for novel nonvolatile memory and magnetic field sensors elements. [11,12, 13] A large number of works have been carried out on the thin films of these systems, however, a careful attention is require to study on bulk polycrystalline samples. [ 9-14,15,16]

Complex impedance spectroscopy is a versatile, non-destructive and powerful tool to analyze the microstructure-property relationship of an electro-ceramic medium assembles of grains, grain boundaries and electrode-electrolyte interface.[17] This method explains the electrical conduction process taking place in systems on applying ac input signal. The output response of complex plane plot appears in the form of semicircles with different characteristic time, representing the electrical conduction phenomenon, across the interfacial regions of the bulk material. This technique is also helpful for analyzing the dielectric relaxation phenomenon in the ferroelectric ceramic and complex composite materials. [18,19,20,21,22,23,24] An equivalent lumped circuit with series or parallel combination of resistance (R), capacitance (C), inductance (L), constant phase element (CPE) provides the information about change in physical processes occurring inside the electro-ceramics.

In this study, we report the synthesis, dielectric, and electrical process of polycrystalline complex PZTFW ceramics and PZTFW-CFO composite. Effect of magnetic phase CFO on the dielectric and electrical properties of PZTFW has been discussed in the context of FPTT.

## II. Experimental Details

Ferroelectric ceramics with complex compositions of $Pb[(Zr_{0.52}Ti_{0.48})_{0.60}(Fe_{0.67}W_{0.33})_{0.40}]O_3$ (PZTFW) and $[Pb[(Zr_{0.52}Ti_{0.48})_{0.60}(Fe_{0.67}W_{0.33})_{0.40}]O_3]_{0.80}$–$[CoFe_2O_4]_{0.20}$(PZTFW-CFO) were synthesized by solid state reaction technique, with initial high purity ingredients: PbO, $ZrO_2$, $TiO_2$, $Fe_2O_3$, $WO_3$, and $Co_3O_4$ (Sigma aldrich - 99.9%). Extra 10% of PbO was added in total amount of PbO to compensate the lead loss during the high temperature calcination and sintering process of PZTFW. Initial ingredients of PZTFW were mixed mechanically in liquid medium (Isopropyl alcohol) for 2 hrs to get the homogeneous mixing, and annealed at optimized temperature 800 $^o$C for 10 hrs. After annealing, the grinded ceramic powder was mixed with small amount of binder PVA (binder polyvinyl alcohol) for pelletization. For circular pellets were pressed at pressure of 4-6x10$^6$ N/m$^2$ to get circular discs of 13 mm in diameter and 1-1.5 mm in thickness. The pellets of PZTFW sintered at optimized temperature 1200 $^o$C for 4 hrs. Twenty weight percent of CFO was mixed in calcined PZTFW powder to get PZTFW-CFO composites, afterward, sintered at an optimized temperature 1000 $^o$C for 4 hrs in high purity alumina crucible to get nearly 94-96% of theoretical density.

X-ray diffraction (XRD) measurements were performed on sintered ceramic surface with monochromatic source of Cu-$k_\alpha$ radiation (BRUKER D8 ADVANCE) over wide range of Bragg's angle (20$^o$<2θ<80$^o$). Microstructure analysis of sintered pellets was observed by scanning electron microscopy (SEM, Zeiss EVO MA-10) technique. For electrical characterization, smooth flat surface of each pellets were electroded with the high purity silver paint, and dried at 200 $^o$C for 2 hrs to remove chemicals. Temperature dependent dielectric constant, dielectric loss, impedance and various electrical measurements were carried out on this metalized pellets at various frequencies (1 KHz~10 MHz) by using an LCR meter (4200-SCS Analyzer) at oscillating amplitude of 0.5V. Ferroelectric P-E hysteresis loops were measured by using a Radiant Ferroelectric Tester.

### III. Results and Discussion:

#### A. Structural Study:

Figure 1(a) shows the XRD patterns of sintered pellets of PZTFW electro-ceramics and PZTFW-CFO composites, in which, PZTFW peaks were carefully indexed for tetragonal crystal structure with space group *P4mm*, and CFO for cubic spinal structure with space group *Fd3m* utilizing UnitCell Win program. The lattice parameters determined for PZTFW single phase were a = 4.078 Å and c = 4.070 Å and for PZTFW-CFO composite were a = 4.060 Å and c = 4.034 Å, matched nearly with morphotropic phase boundary (MPB) of PZT tetragonal crystal structure (with space group P4mm). [25] A significant decrease in lattice parameters was observed in case of composite; in-plane compressive stain, $\eta_{in} = \frac{(a_p - a_c)}{a_p} X100 = \frac{4.078 - 4.060}{4.078} X100 = 0.44$, was 0.44% and out-of-plane i.e. along c-axis, $\eta_{out} = \frac{(c_p - c_c)}{c_p} X100 = \frac{4.070 - 4.034}{4.070} X100 = 0.88$ was 0.88%, (where, $\eta$, $a_c$, $a_p$, $c_p$, and $c_c$ represent, compressive strain, in-plane lattice constant of perovskite, in-plane lattice constant of composite, out-of-plane lattice constant of perovskite, and out-of-plane lattice constant of composite, respectively). This may be due to large stress produced by CFO in PZTFW-CFO composites. Interestingly, CFO also acts as purifier in composites; a small amount of pyrochlore/impurity (< 0.1%) disappeared in PZTFW-CFO composite. To identify the complex pyrochlore/impurity phase with more than six elements in a system is not easy and judicious task to conclude the exact impurity phase. The peak around 28.9° generally referred as $A_2B_2O_7$ and $A_2B_2O_{6-\delta}$ systems, any of the B-site elements, including A-site lead ($Pb^{+4}$) may seat on the B-site of impurity phase. [26] A thorough and systematic study is required to identify the pyrochlore phase. It indicates that pyrochlore phase is converted to the perovskite phase during the growth of composites. A distinct and clear CFO spinal phase Bragg's angle can be seen at 35.36, 36.24 and 62.14 (degree) in composite suggests the formation of two phases without any intermixing. Figure 1(b, c, d) shows shifting and broadening of the PZTFW peaks (111), (211) and (220) toward higher Braggs angle in PZTFW-CFO composites concludes the occurrence of strain as well as change in particle size with the influence of CFO in PZTFW matrix, respectively.

**B. Microstructural Study:**

Large area SEM images of PZTFW single phase and PZTFW-CFO composite are shown in figure 2 (a) and figure 2 (b), respectively. Fig. 2(b) clearly shows that both the phases are distinct without any mixed phases. Secondary electron micrograph images and the elemental analysis suggests that the initial elemental compositions matched with the final compositions of both phases within the limit of experimental errors (+/- 10%). In order to support the elemental compositions, Energy Dispersive X-ray (EDAX) plots of each phase are given in figure 3 (a-c). Elemental analysis of pure PZTFW (Fig. 3(a)), in PZTFW-CFO composite (Fig. 3(b)), and CFO phase in PZTFW-CFO (Fig. 3(c)) composite suggests the presence of appropriate stoichiometric and all the desired elements in single phase and composite medium. In composite, PZTFW and CFO phase show little presence of CFO and PZTFW elements, respectively, it may be due to large penetration depth of energetic electrons and the divergence detection of the emergent x-ray from different energy levels from different compositions. Note that it would be unwise to discuss the exact compositions of the composites since it all depends on the probe area, composites grain distributions, overlapping between two grains, and sometime similar energy level for two different types of investigated elements. Fine spherical PZTFW grains can be seen in both systems; however, in addition of CFO phases slightly reduced the grain sizes of PZTFW, which is well supported by the XRD data. CFO phase is distinct and inhomogenously distributed in the PZTFW matrix within 80% of the weight ratio. SEM images also indicate that grains are densely packed without the presence of voids, cracks, and porosity. Both the phases and their boundaries are well defined and linked. Grains of unequal size are unevenly distributed throughout the sample with average grain size ~ 0.5-2 μm. CFO grains are of needle shape with an average size of ~ 200-300 nm, mostly surrounding the PZTFW grains.

## C. Dielectric Spectroscopy:

PZTFW is a high dielectric constant material and its dielectric constant is much larger than CFO, it would be hard to distinguish the bulk dielectric constant of both the phases. Figure 4 (a,b) & (c,d) show the temperature dependent dielectric constant ($\varepsilon$) and tangent loss (tan $\delta$) of PZTFW and PZTFW+CFO between 5 kHz to 1 MHz, respectively. It indicates second order ferroelectric phase transition at 165 (+/-5) °C, with an unknown kink near 230 °C for lower probe frequency, latter phase transition may occurs due to high conductivity at elevated temperatures. Tangent loss also shows a kink just below the FPTT, which further increases with increase in temperature. PZTFW+CFO composite also illustrates the similar ferroelectric phase transition temperature (FPTT), however, for low probe frequency, the nature of FPTT is diffused-type. High probe frequencies (>50 kHz) both the systems show low dielectric loss. The composite shows step-type temperature dependent dielectric above the FPTT due to low loss. The composite also demonstrates high dielectric constant and high dielectric dispersion below the FPTT compare to the PZTFW single phase, however, conductive nature of CFO plays vital role in enhancement of dielectric properties. Both systems show low dielectric constant below FPTT and very high dielectric loss above 250 °C. Similar dielectric behavior was observed in PZT+CFO multilayer thin films. [27] The nature of FPT is quite different in composite due to addition of semiconducting CFO into the PZTFW matrix. The maximum dielectric constant in PZTFW occurred near FPTT is ~ 2900, that is three times smaller than the PZTFW-CFO composite. The change in dielectric response as function frequency is significant in both cases because of broad dielectric relaxation times and non-Debye type of dielectric relaxation, [28] these observation will be discuss in the section of conductivity and modulus spectroscopy.

**D. Complex Impedance analysis:**

For the analysis of impedance data usually appropriate choice of equivalent circuits and probe frequencies are used for realistic representation of the electrical properties of the material. An ideal equivalent circuit requires explaining the electrical signals from the material that comprises resistance (R), inductance (L), capacitance (C), and constant phase element (CPE). There are several physical models such as Brick Layer model, effective medium theory model, Maxwell-Wagner/Hashin Shtrikman model, Zuzovsky/Brenner model etc. [ 29,30,31,32,33,34] commonly used for getting fit with experimental data. The experimental observations of result can be explained based on real (Z′) and imaginary (Z″) part of impedance.

$$Z^* = Z' - jZ'' \quad \ldots \ldots \quad (1)$$

$$Z^* = Z\cos\theta + jZ\sin\theta \quad \ldots \ldots \quad (2)$$

For electrical modulus data analysis, the complex impedance can be converted to the complex modulus, latter is very effective to analyze the low capacitance of the systems:

$$M^* = i\omega C_0 Z^* = i\omega C_0 (Z' - iZ'') = M' + iM'' \quad \ldots \ldots \quad (3)$$

via $M' = \omega C_0 Z''$ and $M'' = \omega C_0 Z'$ $\ldots \ldots \quad (4)$

$C_0 = \varepsilon_0 A/d$ is the vacuum capacitance of the measuring cell and electrodes with and air gap of the sample thickness, where $\varepsilon_0$ is the permittivity of free space (8.854 x$10^{-12}$ F/m) $d$ is the sample thickness and $A$ is the cross-sectional area of the electrode deposited on the sample; $\omega$ is the angular frequency. Experimental Nyquist plots and their model fittings for both the systems are given in the figure 5 (a-e). A single semicircle arcs are observed for both systems over 1 kHz to 10 MHz (to check the bulk properties), however the bulk resistance of PZTFW-CFO is several orders in magnitude less than the PZTFW, depending on the temperatures and probe frequencies. The physical model used to explain the observed experimental result is based on the parallel combination of resistance, capacitance and constant phase element with series combination of contact resistance. Fitting were carried out using Zview 2 software and the related equivalent circuit is given in figure 5 (e) and all the fitted parameters are given in Table.1 & Table 2. Both systems, PZTFW and PZTFW-CFO indicates partial semicircle below 100°C and 150 °C (below FPTT), suggest highly resistive in nature and contribution to bulk in ferroelectric properties. As temperature goes on increasing, 200-350 °C, it becomes nearly semicircle but not centered on real axis. [35] This indicates that while increasing the temperature the bulk resistance decreases drastically with shift in characteristic relaxation times toward higher frequency side. The CPE is defined in terms of impedance as follows: $1/Z^*_{CPE} = A_0(j\omega)^n$, where n= 0 the lump circuit will behaves like resistive state, and n= 1 for capacitive state (two limiting cases). The variation of critical CPE exponent as function of temperature is shown in figure 6 which illustrates that the CPE acts as capacitive element for PZTFW-CFO near and above the FPTT and decreases rapidly with increase in temperature and becomes more resistive, however, it behaves like a capacitive element near the FPTT in PZTFW ceramics, below and above is more resistive. The CPE exponents support the FPTT. The diameter of semicircle in Nyquist plot gives electrical resistivity at specific temperature and maximum amplitude value corresponds to relaxation frequency. The nature of Nyquist plots is same for both the systems; however, there is a huge difference in resistance which is mainly due to semiconducting behavior of CFO in PZTFW-CFO matrix at elevated temperatures. Therefore rise in temperature leads to conductivity and lowering in impedance values.

A large variation of imaginary part of impedance (Z″) values and relaxation times as function of frequency was observed from temperature 150 °C to 350 °C. With increase in temperature, impedance value decreases several order of magnitudes in the vicinity of FPTT which further merger at higher frequency as shown in figure 7 (a-b). Interestingly, a discontinuity in dielectric relaxation times and their activation energy was observed near the FPTT for both the systems as can be seen in the figure 8 (a) & (b), respectively. In both cases, relaxation time decreases with increase in temperatures which is very fast after FPTT, it also indicates that faster movement of mobile charge carriers responsible for the long range conductivity. The relaxation time of PZTFW-CFO shows an order small magnitude compare to PZTFW over wide range of temperature. This effect deduces that addition of CFO in PZTFW matrix lower its resistivity and generates larger number of mobile charges. The relaxation time ($\tau$) resulting from the Z″ fits, using equation $\omega_{max}\tau=1$, for above and below FPTT against the inverse temperature. We found that the relaxation time follows Arrhenius equation:

$$\tau = \tau_0 \exp\left(\frac{-E_a}{k_B T}\right)$$

and its activation energies calculated from the slopes of the fitted straight lines support the majority of oxygen ion conduction in both the system. [23] Both systems show that the mobile charge carriers follow two distinguished activation energy regions with different energy values. These two activation energy is related to activation energy of the bulk PZTFW and PZTFW-CFO systems before and after the ferroelectric phase transition temperature (FPTT). The slope of the relaxation time, which provides the activation energy of charge carriers, unexpectedly changed after phase transition. We observed that mobile charge carriers move with different energy before and after FPTT and have different magnitudes in single phase and composite system.

**E. Modulus Formalism:**

Complex modulus formalism has its own special advantage to find out the small capacitance, which may arises due to grain boundaries, interfaces across different boundaries/electrodes, or from the small capacitance by other phases in composites. The expression and process for conversion for electrical modulus M′ and M″ are given by equation (3) and (4). Using equation (3) and (4), frequency dependent complex modulus formalism plotted at various temperatures is shown in figure 9 (a-d). Figure 9 (a) & (b) show variation of imaginary part of electrical modulus (M″) of PZTFW and PZTFW-CFO as function of frequency at different temperature, respectively. It can be easily noticed that only single dielectric relaxation peak was observed for PZTFW where as a strong signature of two relaxation peaks at various frequencies for PZTFW-CFO below 250 °C which shadowed beyond the probe frequency at elevated temperatures. With the help of master modulus spectra, it is possible to distinguish the effect of both phases (even CFO has very low capacitance compare to PZFW, not visible in Nyquist plots) to the overall electrical and dielectric properties of composite. CFO shows very fast relaxation time (1.6e-7 s at 150°C) below the FPTT which reached at 3.34e-8 s at 250 °C plays important role in the step-type dielectric behavior above FPTT, which can be seen from the dielectric spectra of PZTFW-CFO. After long time, an important experimental artifact is observed with the help of Modulus formalism which shows a spiral M″ vs M′ behavior above 4 MHz probe frequency that cannot be explain by any physical electrical circuits. [36] Shadowed area illustrates the spiral-type of modulus spectra. In case of PZTFW-CFO composites, there was asymmetrical arc, with step-like broad M″ vs M′ spectra was observed in this composites (as shown in figure 9 (d)). It indicates that CFO significantly affect the bulk capacitance of composite due to their large long range conductivity and fast relaxation times even near the FPTT. The shift in modulus peaks toward higher frequency gives small time constant as per relation $\omega_{max}\tau = 1$. This arising of peak at higher temperature, used to calculate the relaxation time provides that activation energy of different type of mobile charge carriers responsible for conduction process. [37,38,39]

**F. AC conductivity study:**

Frequency dependent conductivity ($\sigma_{ac}$) for PZTFW and PZTFW-CFO was calculated from the dielectric data by using empirical relation:

$$\sigma_{ac} = \omega\varepsilon_r\varepsilon_0 tan\delta \quad \ldots\ldots\ldots\ldots\ldots\ldots\ldots\ldots\ldots \quad (5)$$

where $\sigma_{ac}$ is total electrical conductivity, $\omega$ is angular frequency, $\varepsilon_r$ relative permittivity, $\varepsilon_o$ free space permittivity, $tan\delta$ is dissipation factor. [40] The ac conductivity of PZTFW and PZTFW-CFO was measured as function of frequency at different temperature as shown in figure 10 (a & b), respectively. Jonscher power law relation is an ideal model to fit the experimental conductivity data and to explain the physical properties; however, due to several type of conductivity contribution, this model does not fit for present data. Modified power law (jump relaxation model (JRM)) has been employed to explain the frequency dependent conductivity of the PZTFW and PZTFW-CFO composites, the experimental data has been well fitted to a double power law. [41, 42]

$$\sigma_{ac}(\omega) = \sigma_{dc}(0) + A_1\omega^{n_1} + A_2\omega^{n_2} \quad\ldots\ldots\ldots\ldots\ldots\ldots\ldots\ldots\ldots\ldots\ldots\ldots\ldots\ldots\ldots\ldots\ldots\ldots\ldots\ldots\ldots\ldots\ldots\ldots\ldots\ldots\ldots\ldots\ldots\ldots\ldots\ldots\ldots(6)$$

The term $\sigma_{dc}(0)$ corresponds to the long-range translation hopping gives the DC conductivity of the systems. The second factor $A_1\omega^{n_1}$ is used for the short range translational hopping motion, where the exponent $0 < n_1 < 1$ characterizes the low frequency region. The third factor $A_2\omega^{n_2}$ is associated to a localized or reorientational hopping motion, having exponent $1 < n_2 < 2$.

The ac conductivity graphs of PZTFW and PZTFW-CFO possess following features; i) at room temperature and low frequency, the ac conductivity is almost frequency dependent, ii) it fits well with double exponential power law in the vicinity of FPTT, iii) frequency independent conductivity region increases with increase in temperatures, iv) value of both the exponents are well within the theoretical limits (given in the Table 3 (for PZTFW) and Table 4 (for PZTFW-CFO)), v) an enhancement of 4-5 order of frequency independent conductivity with increase in temperature from 150 °C to 350 °C, however composite shows an order higher magnitude, and vi) at high frequency ($>10^4$ -$10^5$ Hz), two competing relaxation process may be visualized.

The JRM depends on the hopping mechanism of charge carrier ions from its equilibrium position, when ions get a sufficient amount of energy (e.g. thermal energy). Initially these ions relax in localized region but when it gets sufficient amount of energy its configuration will not remain in thermal equilibrium with its surrounding. In order to stabilize or to get in equilibrium at new position another ions have to displace from its equilibrium state. In some cases, ion can also jump back and forth (i.e. unsuccessful hoping) in order to partially relax from the initial state. The conductivity in low frequency region (< 5 kHz) is associated with successful long range translation motion of mobile ions. In mid frequency range ( < 5 MHz), the rate of successful hoping lowers due to two competing relaxation process takes place such as, unsuccessful forward-backward-forward hoping and unlocalized hoping ( remain stay in new position). A dispersive conductivity plateau was observed due to the increase in the ratio of successful to unsuccessful hoping at higher frequency region (> 5 MHz).

**G. Polarization as a function of electric field**:

Polarization versus electric field (P-E) experiment was carried out to establish the ferroelectric nature of both the systems, as shown in the figure 11 (inset shows the PZTFW-CFO P-E loop). Unpoled PZTFW sample shows the slim ferroelectric hysteresis at room temperature with saturation polarization of 8 μC/cm$^2$. Small defects in PZTFW crystal may pinch the hysteresis at the zero applied fields. The coercive field was ~ 4 kV/cm with maximum application of external electric field 30 kV/cm. Bulk polycrystalline samples are difficult to sustain higher applied electric field. A small opening in hysteresis indicates the possible leakage during the application of applied field. P-E hysteresis of PZTFW-CFO (inset fig. 11) composite shows low saturation polarization compare to PZTFW, and its nature is some extent more fatty which may be due to semiconducting nature of CFO in PZTFW matrix.

## IV. Conclusion

In conclusion, we have successfully synthesized the ferroelectric PZTFW and PZTFW-CFO composite having FPTT near 165 (+/- 5) °C. A step-like dielectric behavior near FPTT was observed for composite. XRD data indicates that incorporation of CFO in PZTFW matrix purify the PZTFW phase. High dense grains, negligible porosity, and two distinct phases (for composite) are observed from the SEM images. P-E hysteresis confirms the ferroelectric nature of both systems. Impedance spectra modeled with the parallel combination of grains capacitance,

CPE, and resistance, fit well with the experimental data in the vicinity of FPTT. Bulk resistive contribution of CFO is hard to distinguish form the PZTFW resistance whereas master modulus spectra are able to differentiate the low bulk capacitance of CFO. A unique spiral modulus behavior was observed at high frequency which may be due to possible experimental artifacts due for spring inductance (attached with probe of impedance analyzer). The magnitude of ac conductivity of PZTFW in frequency independent region is at least an order lower than the composite. A system with higher degree of disorderness in lattice or in the matrix follows the double exponential power laws, which is also valid for the present investigation.

Table 1. Fitted parameters of PZTFW electro-ceramics obtained after the utilization of model equivalent circuit in Nyquist plots (fig. 5)

| Temperature (°C) | $R_s$ (Ω) | $R_{bulk}$ (Ω) | $C_{bulk}$ (F) | Constant Phase Element Parameters | |
|---|---|---|---|---|---|
| | | | | $A_0(\omega)$ | n |
| 200 | 10 | 119360 | 1.64E-10 | 6.93E-8 | 0.49 |
| 250 | 28 | 11920 | 5.3E-11 | 2.64E-9 | 0.77 |
| 270 | 19 | 4920 | 2.20E-11 | 9.96E-10 | 0.85 |
| 300 | 20.32 | 1888 | 3.59E-12 | 5.66E-10 | 0.89 |

Table 2. Fitted parameters of PZTFW-CFO composite obtained after the utilization of model equivalent circuit in Nyquist plots (fig.5)

| Temperature (°C) | $R_s$ (Ω) | $R_{bulk}$ (Ω) | $C_{bulk}$ (F) | Constant Phase Element Parameters | |
|---|---|---|---|---|---|
| | | | | $A_0(\omega)$ | n |
| 200 | 5.69 | 13328 | 3.7E-11 | 1.72E-8 | 0.74 |
| 250 | 41.4 | 1541 | 5.70E-11 | 1.66E-8 | 0.70 |
| 270 | 52.36 | 730.6 | 1.008E-10 | 3.77E-8 | 0.65 |
| 300 | 62.03 | 323 | 1.57e-10 | 1.33E-7 | 0.57 |

Table 3. Fitted ac conductivity parameters of PZTFW ceramics acquired after utilization of double exponential power law.

| Temperature (°C) | $\sigma_{dc}$ | $A_1$ | $A_2$ | $n_1$ | $n_2$ |
|---|---|---|---|---|---|
| 200 | 4.6e-6 | 1.03e-8 | 4.49e-20 | 0.56 | 1.87 |
| 250 | 4.1e-5 | 3.6e-8 | 4.49e-18 | 0.48 | 1.79 |
| 300 | 2.87e-4 | 1.1e-7 | 2.07e-12 | 0.39 | 1.22 |
| 350 | 1.1e-3 | 4.22e-11 | 9.7e-18 | 0.99 | 1.97 |

Table 4. Fitted ac conductivity parameters of PZTFW-CFO composites acquired after utilization of double exponential power law.

| Temperature (°C) | $\sigma_{dc}$ | $A_1$ | $A_2$ | $n_1$ | $n_2$ |
|---|---|---|---|---|---|
| 200 | 3.3e-5 | 2.2e-8 | 6.7e-18 | 0.59 | 1.8 |
| 250 | 3.3e-4 | 5.3e-7 | 5.5e-15 | 0.43 | 1.74 |
| 300 | 1.4e-3 | 2.4e-6 | 8.5e-14 | 0.34 | 1.5 |
| 350 | 3.2e-3 | 6.8e-8 | 2.7e-17 | 0.62 | 1.98 |

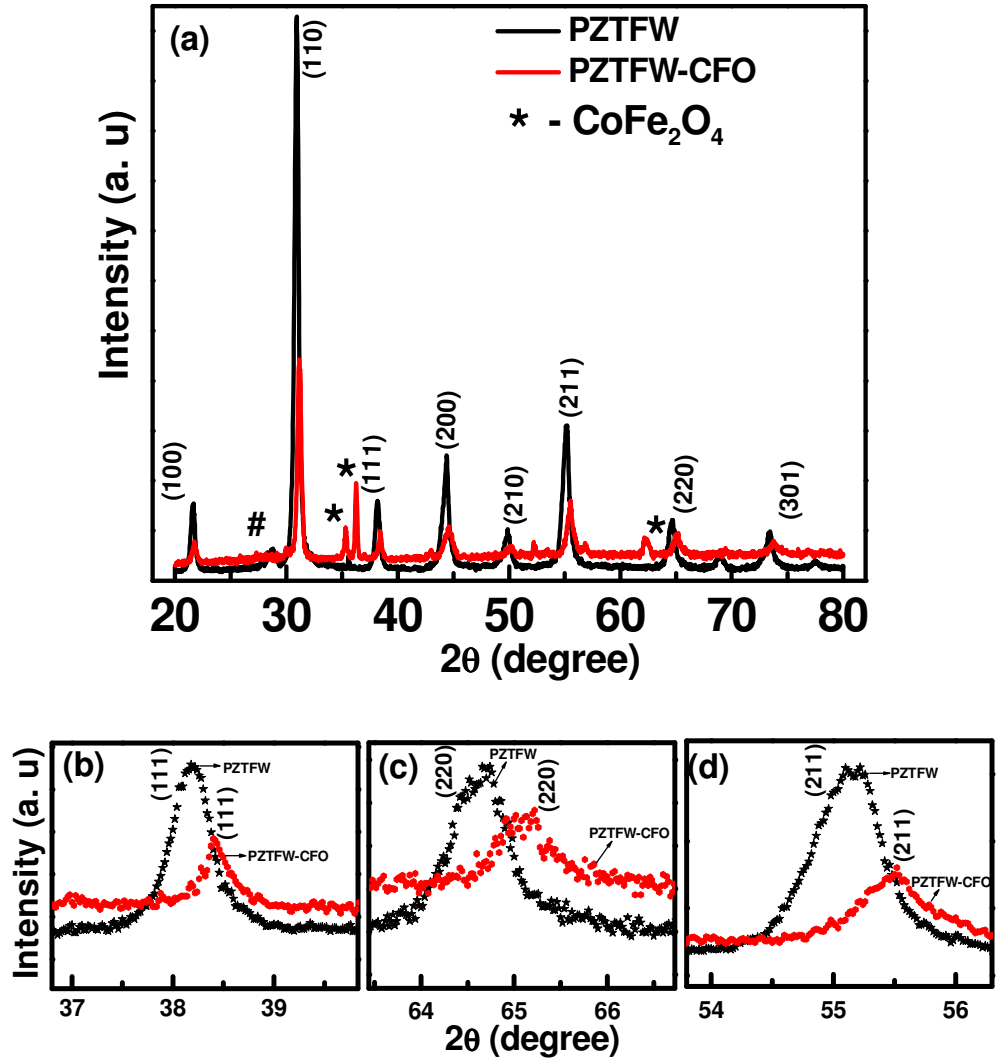

Figure 1. (a) XRD patterns of PZTFW and PZTFW-CFO composite. (*) indicates the patterns related to CFO phase, (b) magnified view of (111), (c) magnified view of (220), (d) magnified view of (211) plane, respectively, and its shift to higher Braggs angle side due to in built compressive strain, plane and its shift to higher Braggs angle side due to in-built compressive strain.

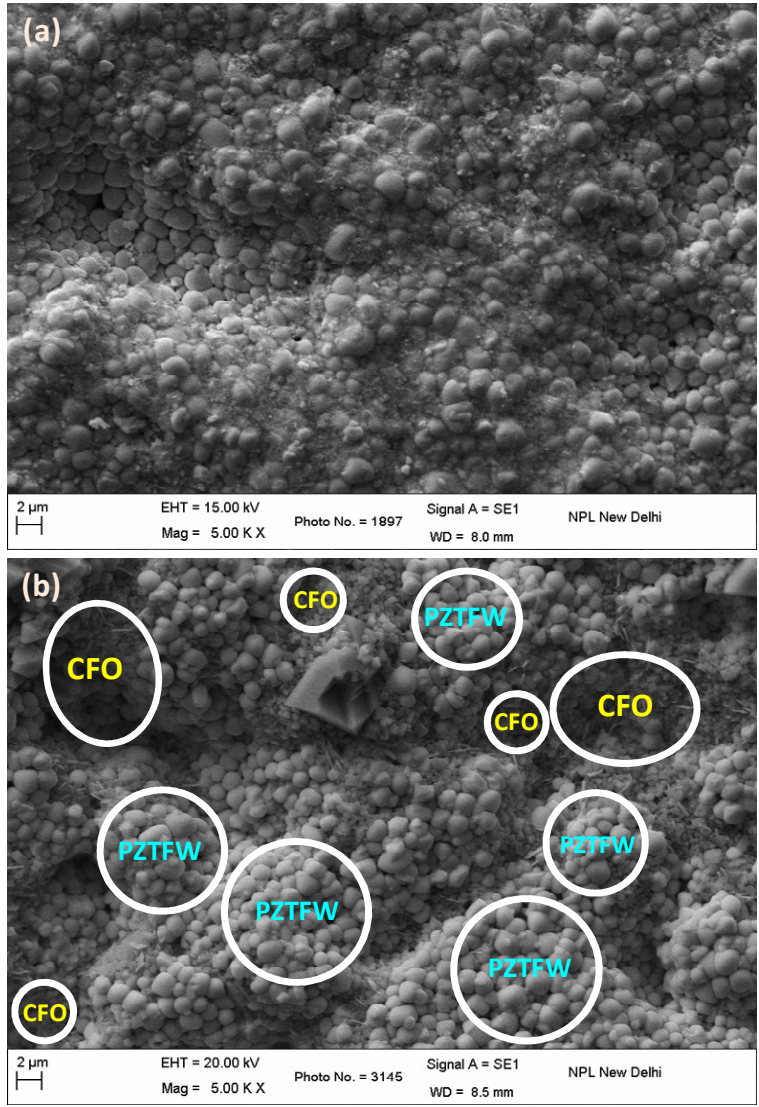

Figure 2 (a) SEM images of (a) PZTFW, (b) PZTFW-CFO, circles indicate the presence of both the phases with densely packed grains.

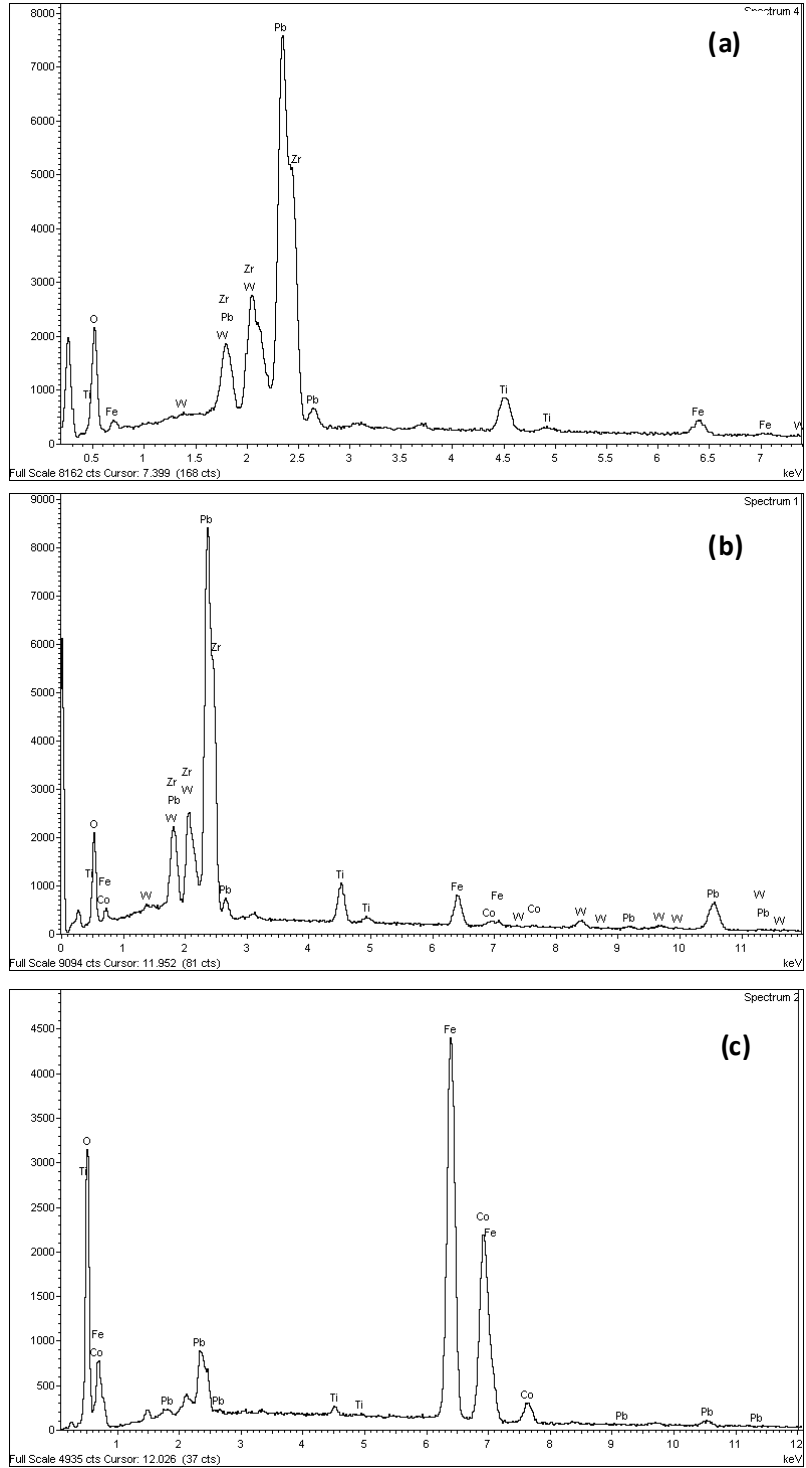

Fig.3 Energy Dispersive X-ray elemental analysis (EDAX) plot of (a) pure PZTFW, (b) PZTFW phase in PZTFW-CFO composite, and (c) CFO phase in PZTFW-CFO composite.

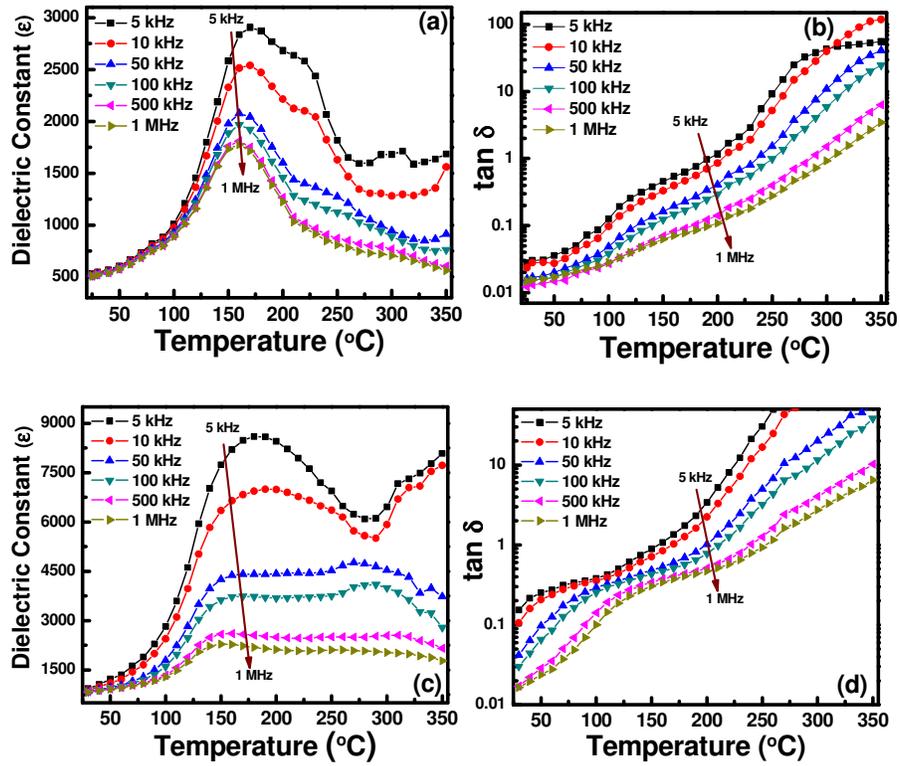

Figure 4 Temperature dependent dielectric and tangent loss spectra of (a-b) PZTFW, (c-d) PZTFW-CFO, respecively, over wide range of frequencies ( 5 kHz – 10 MHz).

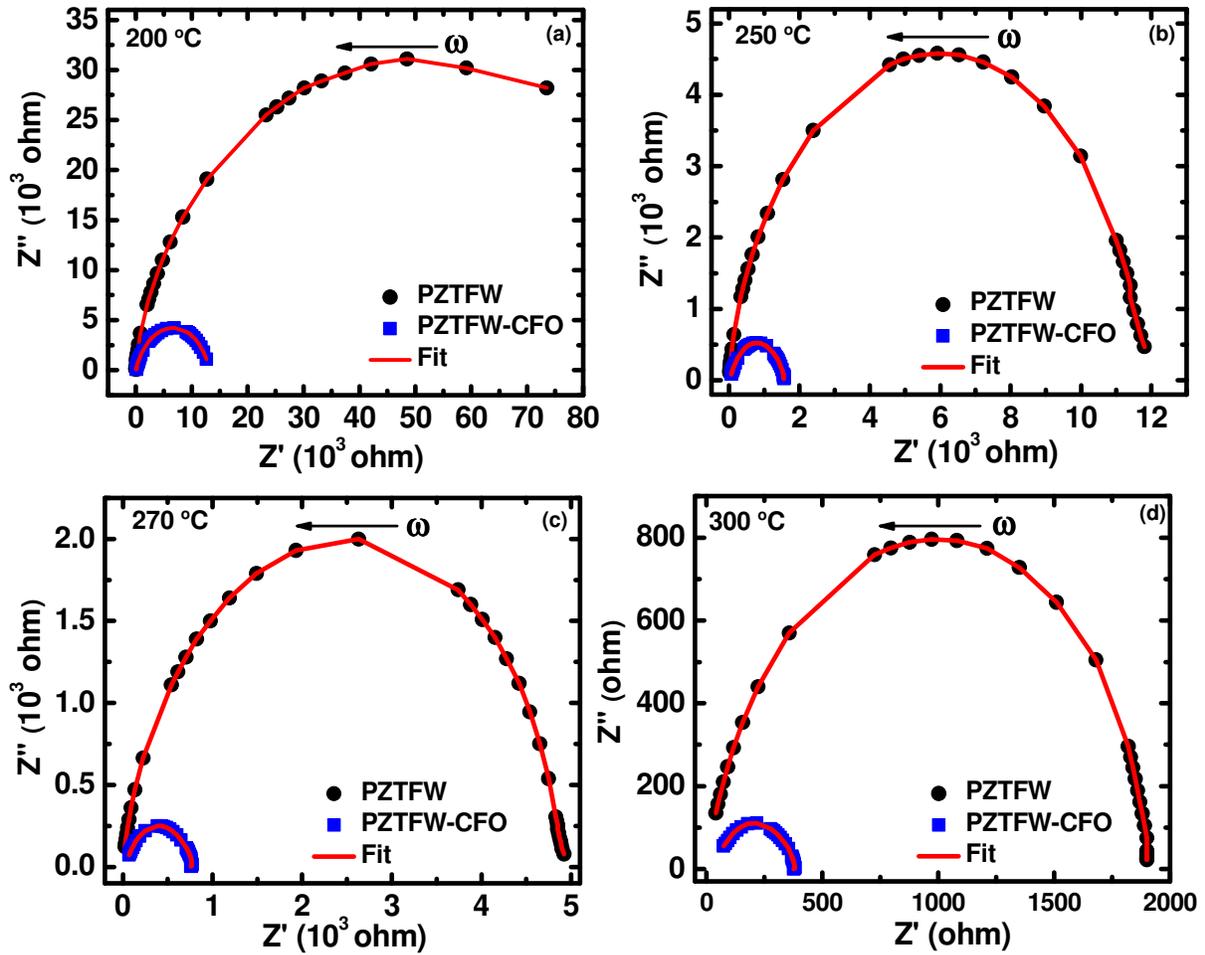

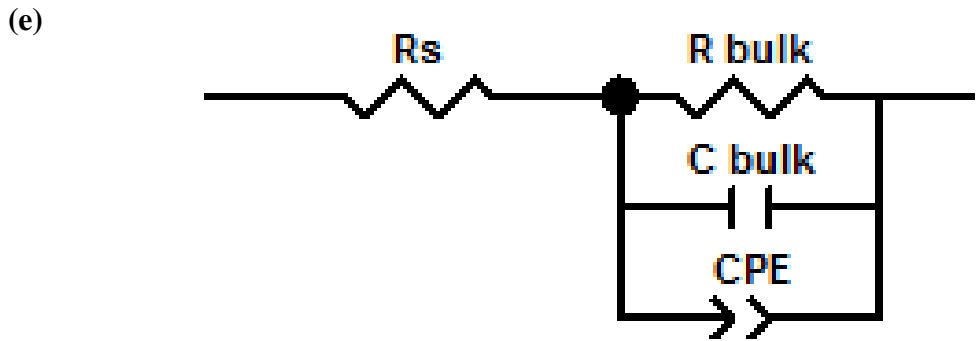

Figure 5 Nyquist plots for both systems as function of frequency at various temperatures of (a) 200 °C, (b) 250 °C, (c) 270 °C, (d) 300 °C, and (e) Model equivalent circuit used for the fitting of experiment Nyquist plot (Fig 5 (a-d))

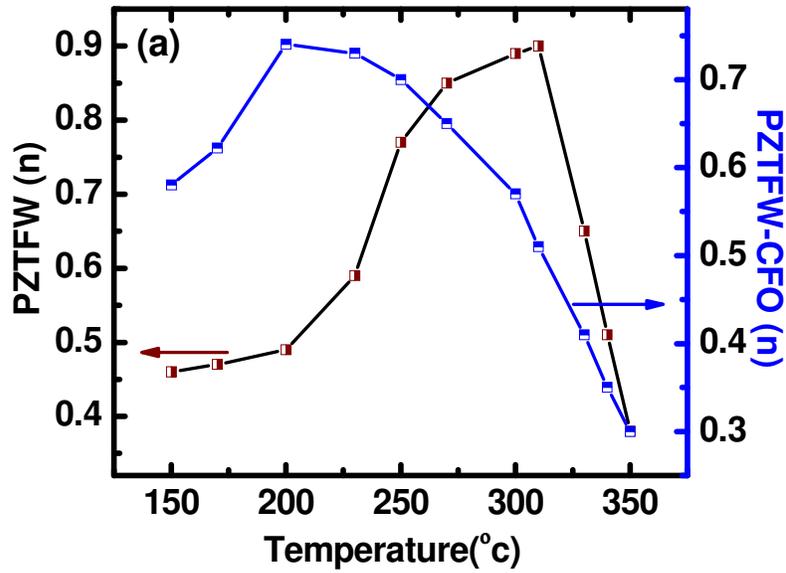

Figure 6 Variation of constant phase element exponent (n) (CPE= $1/Z^*_{CPE} = A_0(j\omega)^n$) as function of temperature near the FPTT.

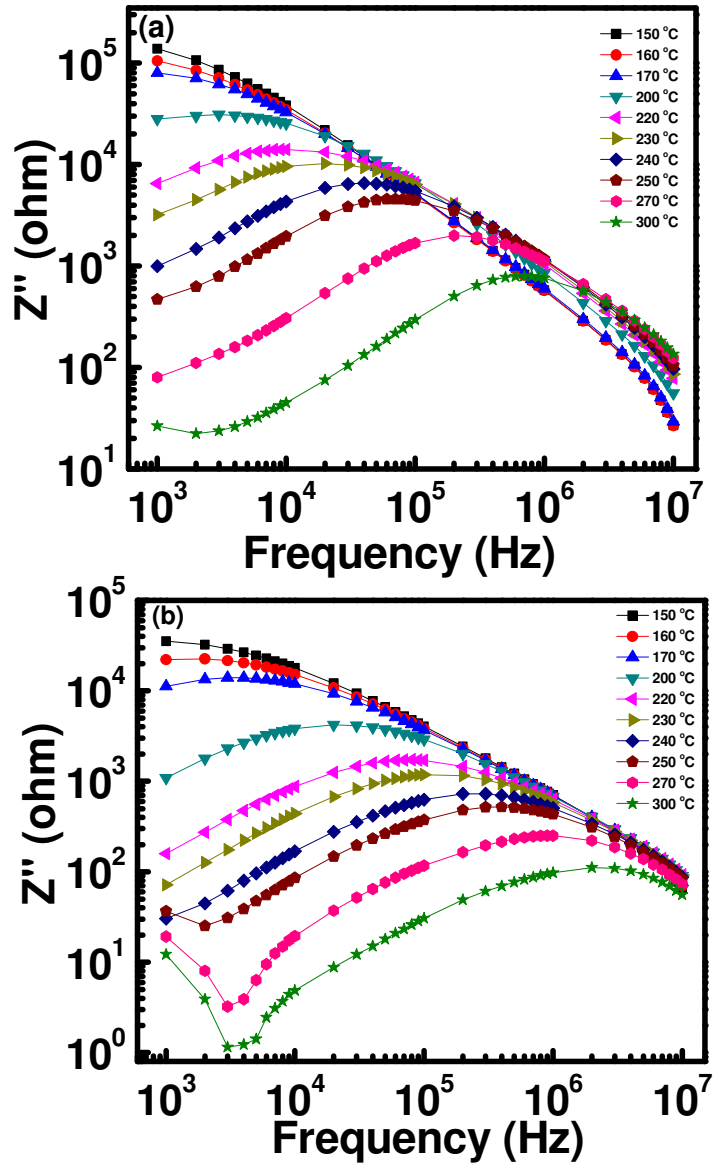

Figure 7 Imaginary part of impedance as function of frequency at various temperatures of (a) PZTFW, (b) PZTFW-CFO, respecively.

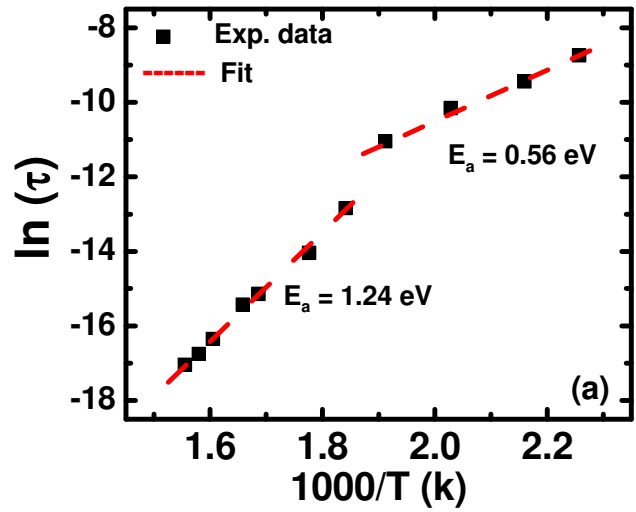

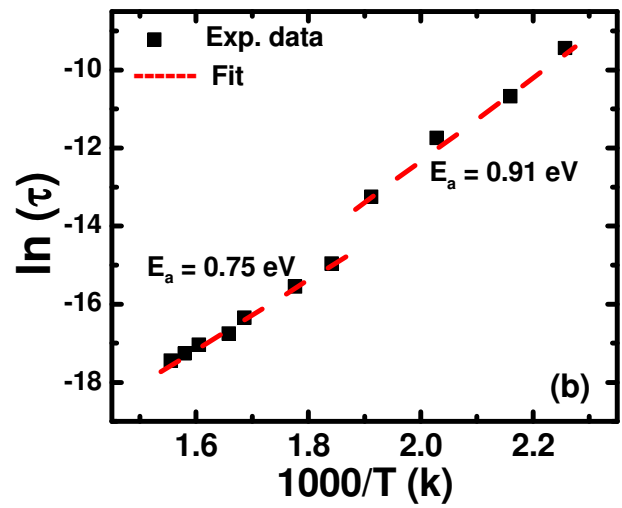

Figure 8 Relaxation time of (a) PZTFW, (b) PZTFW-CFO and its Arrhenius fitting as function of inverse temperatures, respectively.

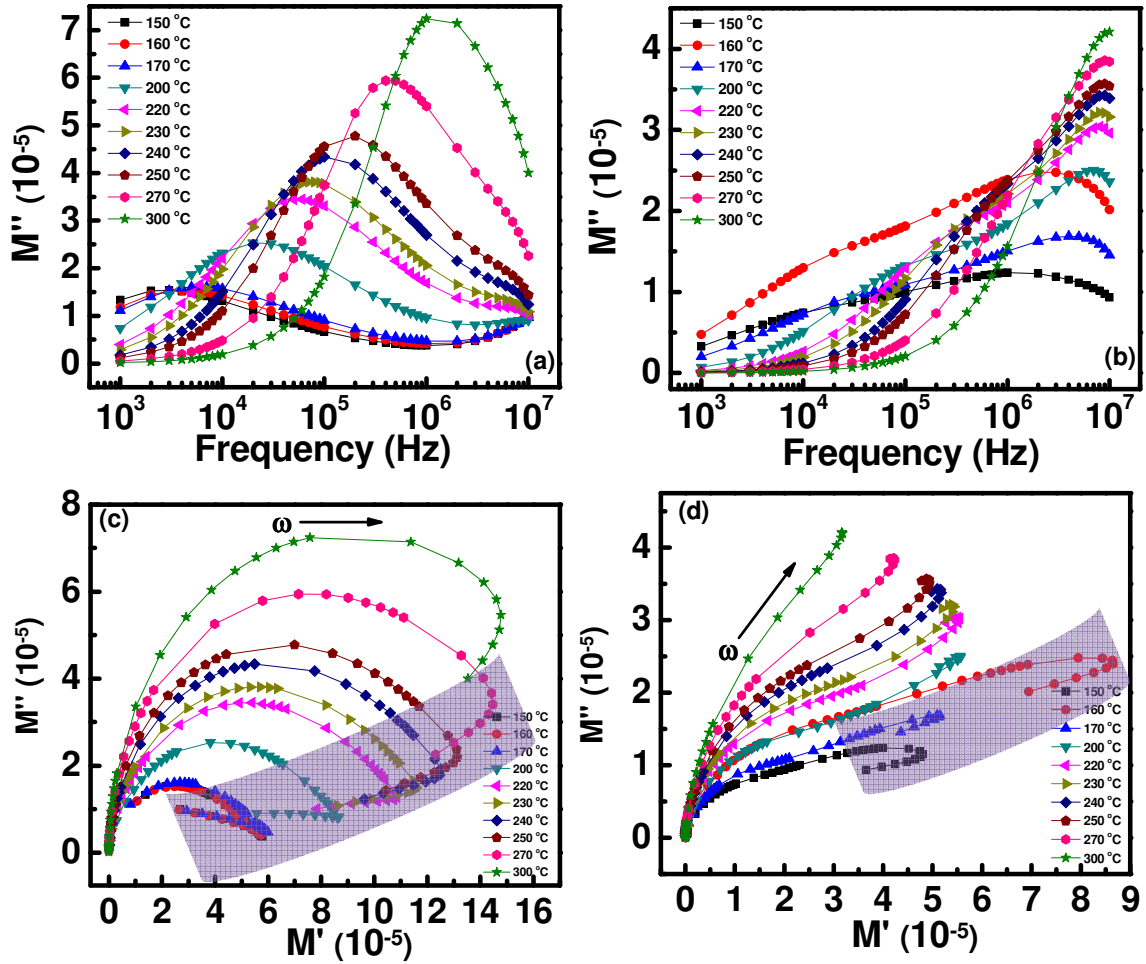

Figure 9 Master modulus spectra as function of frequency at various temperatures , (a, c) PZTFW, and (b, d) PZTFW-CFO.

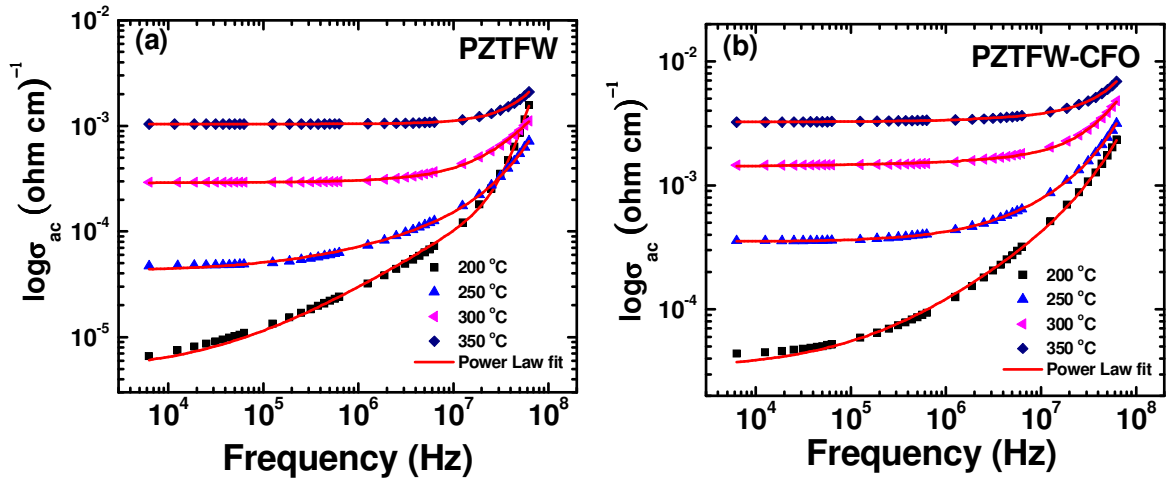

Figure 10 ac condiuctivity plots for both systems (a) PZTFW, (b) PZTFW-CFO as function of frequency at various temperatures (solid line shows modified power law fit).

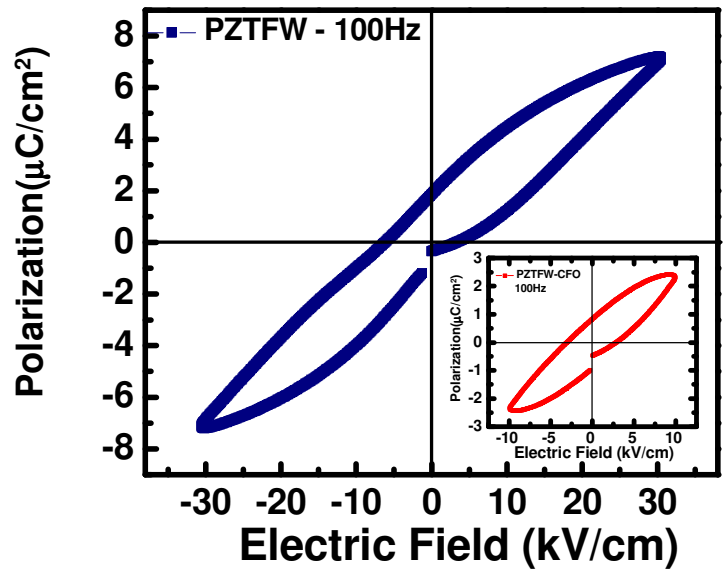

Figure 11 P-E hysteresis plots PZTFW, and PZTFW-CFO (inset) at 100 Hz.